\documentclass[twocolumn,twoside,slac_two]{revtex4}
\usepackage{graphicx}
\usepackage{fancyhdr}
\begin{document}

\title{$\gamma$ measurements at LHCb}

%

\author{J. Nardulli\\{\it{On behalf of the LHCb collaboration}}}
\affiliation{Science and Technology Facility Council, Rutherford Appleton Laboratory, Didcot, UK}

\begin{abstract}
The LHCb collaboration has studied various promising ways to determine
the  Unitarity Triangle angle $\gamma$.
Three complementary methods will be considered.
The potential of the $B \to DK^{(*)}$
decays has been studied by employing the
combined Gronau-London-Wyler (GLW) and the Atwood-Dunietz-Soni (ADS) methods,
making use of a large sample of simulated data.
$\gamma$ can also be extracted
with a time-dependent analysis of $B_s \to D_s K$ decays,
provided that the $B_s$ mixing phase is measured independently.
In addition, the combined measurement of the $B^0 \to \pi^+ \pi^-$ and $B^0_s \to K^+K^-$
time-dependent CP asymmetries
allows the determination of $\gamma$,
up to U-spin flavour symmetry breaking corrections. 
For each method the expected sensitivities to the angle $\gamma$ are presented.
\end{abstract}

\maketitle

\thispagestyle{fancy}


\section{Introduction}
\begin{figure}[h]
  \centering
  \includegraphics[width=0.45\textwidth]{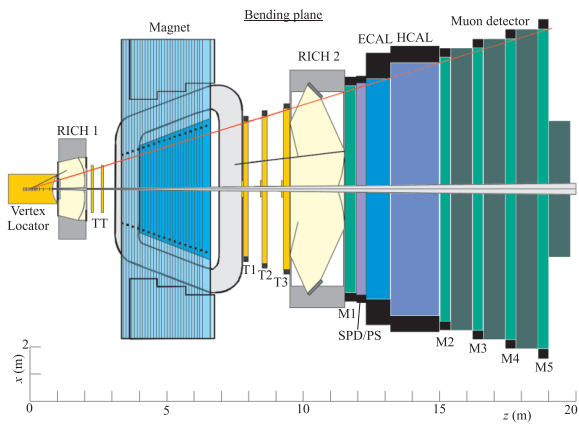}
  \caption{The LHCb setup with the different sub-detectors
    shown in the horizontal (bending) plane.}\label{fig:lhcb}
\end{figure}

LHCb aims to study CP violation and rare $B$-meson decays with high precision, using
the Large Hadron Collider (LHC), where all species of $B$-mesons are produced in
14~$\mathrm{TeV}$ $pp$ collisions~\cite{bib:lhcb1,bib:lhcb2}.
In these events the $b\bar{b}$ pairs
are frequently produced in the same forward (or backward) direction.
The LHCb detector is a single-arm spectrometer with a forward coverage from
10~$\mathrm{mrad}$ to 300~$\mathrm{mrad}$
in the horizontal plane (i.e., the bending plane of the magnet). The acceptance
lies between 10-250~$\mathrm{mrad}$ in the vertical plane (non-bending plane).
The detector layout in the bending plane is shown in Fig.~\ref{fig:lhcb}.

\section{Event Selection}

LHCb will collect large samples of all types of $B$ mesons and baryons. 
These samples will allow the precise measurement of all the three ($\alpha$,
$\beta$ and $\gamma$) angles of the Unitarity triangle and of the $B_s$ mixing phase.
Here, three complementary methods to extract the  Unitarity Triangle angle
$\gamma$ will be considered.

As it will become clear in the next sections, the three complementary methods
considered for the extraction of 
the Unitarity Triangle angle $\gamma$ use different techniques to determine
$\gamma$. However similar approaches are used in the event selection and in the
selection at the trigger level.
In all cases, samples of signal and background events are simulated through
the LHCb apparatus; the selection is then optimized in order to maximize the signal efficiency and 
minimize the background contribution. The main selection criteria used take into
account that the large $B$ mass produces decay products with high $p_T$ and that
the long $B$ lifetime produces tracks with large impact parameter with respect to
the primary vertex. Furthermore particle identification cuts are used for the
$\pi/K$ identification and cuts on the significance of the flight distance of the
$B$ are used in order to have a $B$ detached from the primary vertex. 

All the following results are for an integrated luminosity of 2~fb$^{-1}$, which
corresponds to one nominal year of data taking.

\section{Extracting $\gamma$ from $B \to D K$ decays}

Interfering tree diagrams in the $B^{\pm} \to \tilde{D} K^{\pm}$ decays allow the determination of
the  Unitarity Triangle  angle $\gamma$.  Here, $\tilde{D}$ can be a $D^0$ or a
$\overline{D^0}$ and the $D^0$ and the $\overline{D^0}$ are reconstructed in a
common final state.

In LHCb \cite{bib:guy} this is done by employing the combined Gronau-London-Wyler (GLW) \cite{bib:glw} and the Atwood-Dunietz-Soni
(ADS) \cite{bib:ads} methods. In the first method 
the information of the CP even eigenstates, where a $\tilde{D}$ decays to
$\pi^+\pi^-$ or $K^+K^-$ is used. Note that reconstruction of the CP odd
eigenstates, which include neutral particles, is rather challenging in LHCb and
is not considered in this analysis. The ADS method exploits the
interference between the favoured and double Cabibbo suppressed decay modes of
the neutral D mesons to state such as $K\pi$ or $K\pi\pi\pi$. Using the ADS
method the following equations can be written for the decay rates for the 
neutral $B^0 \to D^0 (\to K\pi) K^{*0}$ channels \cite{bib:kazu} (very similar 
equation can be written for the charged decays, see also \cite{bib:patel}) :
\begin{eqnarray}
\Gamma(B^0 \to (K^+\pi^-)_D K^{*0}) &=& N_{K\pi} (1+(r_B r_D)^2 + 
2r_Br_D \cdot \nonumber \\ & & \cos(\delta_B+\delta_D+\gamma))\;\;\;, \\
\Gamma(B^0 \to (K^-\pi^+)_D K^{*0}) &=& N_{K\pi}(r_B^2+r_D^2+ 
2r_Br_D\cdot \nonumber \\ & & \cos(\delta_B-\delta_D+\gamma))\;\;\;,  \\
\Gamma(\overline{B^0} \to (K^-\pi^+)_D \overline{K^{*0}}) &=& N_{K\pi}(1+(r_B
r_D)^2 + 2r_Br_D\cdot  \nonumber \\ & & \cos(\delta_B+\delta_D-\gamma)) \;\;\;,\\
\Gamma(\overline{B^0} \to (K^+\pi^-)_D \overline{K^{*0}}) &=&
N_{K\pi}(r_B^2+r_D^2+ 2r_Br_D\cdot \nonumber \\ & &cos(\delta_B-\delta_D-\gamma))\;\;\;,
\end{eqnarray}
where 
$$ r_B = \frac{|A(B^0 \to D^0 K^{*0})|}{|A(B^0 \to \overline{D^0} K^{*0})|} $$
is the ratio of the magnitudes between the two amplitudes of the B-decay and
similarly $$ r_D = \frac{|A(D^0 \to K^+\pi^-)|}{|A(\overline{D^0} \to K^+\pi^-)|}\;\;\;.$$
Furthermore $\delta_B$ represents the strong phase difference between the two
B-decays while $\delta_D$ represents the strong phase difference between the two D-decays.
$N_{K\pi}$ gives the overall normalization and represents the total number of
$B^0 \to (K\pi)_D K^{*0}$ events. It can be seen that there are two favoured 
rates (1) and (3) and two suppressed rates (2) and (4). Further information can be added 
by including the decays to CP-eigenstates, such as $\pi^+\pi^-$ or $K^+K^-$ and so:
\begin{eqnarray}
\Gamma(B^0 \to D_{CP} K^{*0}) &=& N_{CP} (1+r_B^2 + \nonumber \\ & & 2r_B\cos(\delta_B+\gamma))\;\;\;,\\
\Gamma(\overline{B^0} \to D_{CP} \overline{K^{*0}}) &=& N_{CP} (1+r_B^2 +
\nonumber \\ & & 2r_B\cos(\delta_B-\gamma))\;\;\;.
\end{eqnarray}
Here, $N_{CP}$ represents the total number of $B^0 \to D_{CP} K^{*0}$
events. Note that $r_D$ is known and is equal to 0.060 $\pm$ 0.003~\cite{bib:pdg}
and $N_{CP}$ can be calculated from $N_{K\pi}$ taking into account the different
efficiencies and branching ratios. Furthermore, for what concerns
$r_B$, for the neutral $B$, values smaller than 0.6 with 95\%  probability have been measured at
Babar \cite{bib:babarneutro}, while for the charged
$B$ the latest world average value is $r_B = 0.10 \pm 0.02$ \cite{bib:ut}.
$\delta_B$ is unknown for the neutral $B$-meson, while its value is fixed at 130$^{\circ}$ for the
charged $B$-meson \cite{bib:babar,bib:belle}. 
$\delta_D$ is assumed to be in the range [-25$^{\circ}$; +25$^{\circ}$] due to a
limit set by the CLEO-c collaboration \cite{bib:cleoc}. In total there are 5 unknowns ($\gamma$, $r_B$, $\delta_B$, $\delta_D$
and $N_{K\pi}$) and 6 observables.

\subsection{Sensitivity to $\gamma$}

As a result of the event selection, the annual yields 
together with the total efficiencies and with the background to signal
ratios are listed in Table~\ref{tab:dkselres} ~\cite{bib:kazu,bib:patel}.
\begin{table}[htbp!]
  \begin{center}
    \vspace{0.5cm}
    \begin{tabular}{|c|c|c|c|} \hline
      Channel & $\epsilon_{tot}$ [\%] & S & $B_{b\bar{b}}/S$  \\ \hline
      $B^0 \to \overline{D^0} (K^+\pi^-) K^{*0} $ & 0.33 & 3350 & $<$ 2.0 \\ 
      $B^0 \to D^0 (K^+\pi^-) K^{*0} $ & 0.33 & 536 & $<$ 12.8 \\
      $B^0 \to D_{CP} (K^+K^-) K^{*0} $ & 0.46 & 474 & $<$ 4.1 \\
      $B^0 \to D_{CP} (\pi^+\pi^-) K^{*0} $ & 0.36 & 134 & $<$ 14 \\\hline\hline
      $B^+ \to \overline{D^0} (K^+\pi^-) K^{+} $ & 0.50 & 28000 & 0.63  \\ 
      $B^+ \to \overline{D^0} (K^-\pi^+) K^{+} $ & 0.50 & 100 & 7.8  \\ 
      $B^+ \to \overline{D^0} (K^+K^-) K^{+} $ & 0.51 & 3000 & 1.2  \\ 
      $B^+ \to \overline{D^0} (\pi^+\pi^-) K^{+} $ & 0.58 & 1000 & 3.6  \\ \hline\hline
    \end{tabular}
   \caption{\label{tab:dkselres}
      Summary of event yield ($S$), 
      experimental efficiency ($\epsilon_{tot}$), and background
      to signal ratios.}
  \end{center}
\end{table}

Both for the charged and neutral $B$-meson decays a standalone Monte Carlo
(MC) simulation was used to generate the event yields and fit the unknown parameters. All the
unknown parameters have been scanned as shown in Table~\ref{tab:scan}. The input value of
$\gamma$ has been fixed at 60$^{\circ}$.
\begin{table}[htbp!]
  \begin{center}
    \begin{tabular}{|c|c|c|}\hline
      Parameter & \multicolumn{2}{|c|}{Scan range} \\\hline
      &  $B^0 \to D^0 K^{*0}$ & $B^{\pm} \to D^0 K^{\pm}$ \\ \hline\hline
      $\delta_B$ & [-180$^{\circ}$ ; +180$^{\circ}$] & fixed at 130$^{\circ}$ \\
      $\delta_D$ & [-25$^{\circ}$ ; +25$^{\circ}$] & [-25$^{\circ}$ ; +25$^{\circ}$] \\
      $r_B$ & [0.0 ; 0.6] & [0.0 ; 0.2] \\ \hline\hline
  \end{tabular}
   \caption{\label{tab:scan}
     Scan ranges of the input parameters of the toy used for the extraction of  
     $\gamma$ for both the charged and the
     neutral $B$-mesons. Note that while $\delta_B$ has been scanned in the full
     range for the neutral $B$-meson, its value was fixed at 130$^{\circ}$ for the
     charged $B$-meson \cite{bib:babar,bib:belle}. 
     $\delta_D$ has been scanned in the range [-25$^{\circ}$; +25$^{\circ}$] due to a
     limit set by the CLEO-c collaboration \cite{bib:cleoc}. For what concerns
     $r_B$, for the neutral $B$, values smaller than 0.6 with 95\%  probability have been measured at
     Babar \cite{bib:babarneutro}, while for the charged
     $B$ the latest world average value is $r_B = 0.10 \pm 0.02$ \cite{bib:ut}.}
  \end{center}
\end{table}
For the charged $B$-meson with one year of data at nominal luminosity
(2fb$^{-1}$), the angle $\gamma$ can be determined with a precision in the range
$8.2^{\circ}-9.6^{\circ}$, depending on the value of the strong phase $\delta_D$ \cite{bib:patel}.
For the neutral $B$ the angle $\gamma$ can be determined with a precision
smaller than $10^{\circ}$, depending on the value of the strong phase $\delta_B$ and
for $r_B$ values bigger than 0.3 \cite{bib:kazu}.

\section{Extracting $\gamma$ from $B^0_s \to D_s K$ decays}

The relations between the $B^0$-meson mass eigenstates $|B_{H, L}\rangle$ and their
flavour eigenstates $|B^0\rangle$ and $\overline{|B^{0}}\rangle$, can be
expressed in terms of linear coefficients $p$ and $q$:
\begin{equation}
  \label{eq:BHL}
  |B_{H,L}\rangle = p |B^0 \rangle \mp q \overline{|B^{0}}\rangle \;\;\;.
\end{equation}

The difference in mass and decay rates are defined as:
\begin{equation}
  \Delta m = m_H - m_L\;\;\;\;,\;\;\;  \Delta \Gamma = \Gamma_H - \Gamma_L\;\;\;.
\end{equation}
The average mass and decay rate are defined as:
\begin{equation}
  m = \frac{m_H + m_L}{2}\;\;\;\;,\;\;\;  \Gamma = \frac{\Gamma_H + \Gamma_L}{2}\;\;\;.
\end{equation}
The decay rate at a time $t$ of an originally produced $|B^0\rangle$ to a final
state $f$ is given by:
\begin{equation}
\Gamma_{B^0 \to f}(t) = |\langle f| T| B^0(t)\rangle|^2\;\;\;,
\end{equation}
where T is the transition matrix element.
The time evolution of the flavour eigenstates  $|B^0\rangle$ and
$|\overline{B^{0}}\rangle$ is then given by the four decay equations:
\begin{eqnarray}
  \label{eq:master}
  \Gamma_{B \to f}(t) &=&
  \left|A_f\right|^2 (1+|\lambda_f|^2) \frac{e^{-\Gamma t}}{2} \cdot
  \nonumber \\ & &
  ( \cosh{\frac{\Delta\Gamma t}{2}} + D_f \sinh{\frac{\Delta\Gamma t}{2}} +
  \nonumber \\ & &
  C_f \cos{\Delta mt} - S_f \sin{\Delta mt} ) \;\;\; , \nonumber \\
  \Gamma_{\overline{B} \to f}(t) &=&
  \left|A_f\right|^2 \left|\frac{p}{q}\right|^2 (1+|\lambda_f|^2)
  \frac{e^{-\Gamma t}}{2} \cdot
  \nonumber \\ & &
  ( \cosh{\frac{\Delta\Gamma t}{2}} + D_f \sinh{\frac{\Delta\Gamma t}{2}}
  \nonumber \\ & &
  - C_f \cos{\Delta mt} + S_f \sin{\Delta mt} )\;\;\;, \nonumber \\
  \Gamma_{\overline{B} \to \overline{f}}(t) &=&
  \left|\overline{A}_{\overline{f}}\right|^2 (1+|\overline{\lambda}_{\overline{f}}|^2) \frac{e^{-\Gamma t}}{2} \cdot
  \nonumber \\ & &
  ( \cosh{\frac{\Delta\Gamma t}{2}} + D_{\overline{f}}
  \sinh{\frac{\Delta\Gamma t}{2}}
  \nonumber \\ & &
  + C_{\overline{f}} \cos{\Delta mt} - S_{\overline{f}} \sin{\Delta mt} ) \;\;\; , \nonumber \\
  \Gamma_{B \to \overline{f}}(t) &=&
  \left|\overline{A}_{\overline{f}}\right|^2 \left|\frac{p}{q}\right|^2 (1+|\overline{\lambda}_{\overline{f}}|^2)
  \frac{e^{-\Gamma t}}{2} \cdot
  \nonumber \\ & &
  ( \cosh{\frac{\Delta\Gamma t}{2}} + D_{\overline{f}}
  \sinh{\frac{\Delta\Gamma t}{2}}
  \nonumber \\ & &
  - C_{\overline{f}} \cos{\Delta mt} + S_{\overline{f}} \sin{\Delta mt})\;\;\;, \nonumber \\
\end{eqnarray}
where
\begin{eqnarray}
  \label{eq:cfdf}
  D_f &=& \frac{2Re{\lambda_f}}{1+|\lambda_f|^2} \;\;\; , \nonumber \\
  C_f &=& \frac{1-|\lambda_f|^2}{1+|\lambda_f|^2} \;\;\; , \nonumber \\
  S_f &=& \frac{2Im{\lambda_f}}{1+|\lambda_f|^2} \;\;\; .
\end{eqnarray}
$A_f$ and $\overline{A}_{\overline{f}}$ are the decay amplitudes (e.g. $A_{f} =
\langle f |T| B^0\rangle$) and 
\begin{equation}
  \label{eq:lambda}
  \lambda_f = \frac{q}{p}\frac{\overline{A}_f}{A_f} \;\;\; , \;\;\;
  \lambda_{\overline{f}} = \frac{q}{p}\frac{\overline{A}_{\overline{f}}}{A_{\overline{f}}} \;\;\; .
\end{equation}

\begin{figure}[htbp!]
  \begin{center}
    \includegraphics[width=0.45\textwidth]{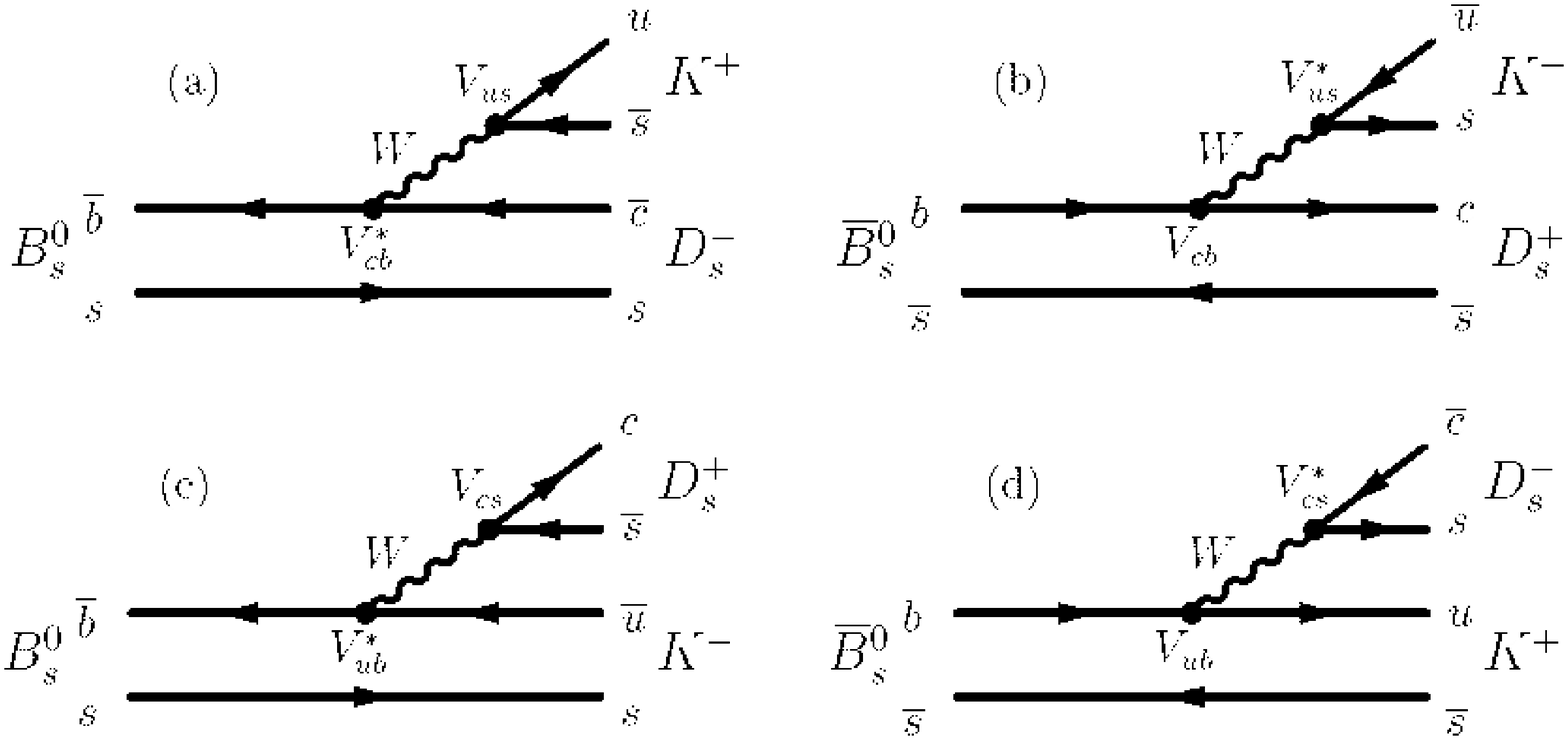}
    \caption{\label{fig:bsds} Feynman diagrams for the $B_{s}^0 \to
    D_s^{-}K^{+}$ (a), $\overline{B_{s}^0} \to D_s^{+}K^{-}$ (b), $B_{s}^0 \to
    D_s^{+}K^{-}$ (c), $\overline{B_{s}^0} \to D_s^{-}K^{+}$ (d) decays.}
  \end{center}
\end{figure}

For the $B_{s}^0 \to D_s^{\mp} K^{\pm}$ decay channels (see Feynman diagrams in
Fig.~\ref{fig:bsds}) a $B^0_s$, as well as a $\overline{B^0_s}$, can decay
directly to $D_s^{-} K^{+}$ or $D_s^{+} K^{-}$. 
In addition, these relations hold for the decay amplitudes:
$|A_f| = |\overline{A}_{\overline{f}}|$ and $|A_{\overline{f}}| =
|\overline{A}_f|$. Assuming $|\frac{q}{p}|=1$, it is possible to write
$|\lambda_f| = |\overline{\lambda}_{\overline{f}}|$. 
The terms $\lambda_f$ and $\overline{\lambda}_{\overline{f}}$
are calculated as
\begin{eqnarray}
  \label{eq:lambdaDsK}
  \lambda_{D^{-}_s K^{+}}  &=& \left(\frac{q}{p}\right)_{B_s}
                         \frac{\overline{A}_{D^{-}_s K^{+}}}{A_{D^{-}_s K^{+}}}
                      = \left(\frac{V_{tb}^*V_{ts}}{V_{tb}V_{ts}^*}\right)
                        \left(\frac{V_{ub}V_{cs}^*}{V_{cb}^*V_{us}}\right) \cdot
                         \nonumber \\
                     & &   \left|\frac{A_2}{A_1}\right| e^{i\Delta_s}
                      = |\lambda_{D^{-}_s K^{+}}|
                        e^{i(\Delta_s - (\gamma+\phi_s)) } \; ,
  \nonumber \\
  \overline{\lambda}_{D^{+}_s K^{-}}  &=& \left(\frac{p}{q}\right)_{B_s}
                         \frac{A_{D^{+}_s K^{-}}}{\overline{A}_{D^{+}_s K^{-}}}
                      = \left(\frac{V_{tb}V_{ts}^*}{V_{tb}^*V_{ts}}\right)
                        \left(\frac{V_{ub}^*V_{cs}}{V_{cb}V_{us}^*}\right) \cdot \nonumber \\
                    & &\left|\frac{A_2}{A_1}\right| e^{i\Delta_s}
                      = |\lambda_{D^{-}_s K^{+}}|
                        e^{i(\Delta_s + \gamma+\phi_s) } \; ,
  \nonumber \\
\end{eqnarray}
where $|A_2/A_1|$ is the ratio of the {\em hadronic} amplitudes, which is
expected to be of order unity, $\Delta_s$ is the {\em strong} phase
difference between $A_1$ and $A_2$ and $\gamma+\phi_s$ is the weak phase. 
The $\phi_s$ angle originates from $B^0_s$ mixing and can be measured directly
using the $B_{s}^0 \to J/\psi \phi$ decay~\cite{bib:lfernandez}.

\subsection{Sensitivity to $\gamma + \phi_s$}

As a result of the event selection, the annual yields together with the total efficiencies and with the background to signal
ratios are listed in Table~\ref{tab:selres}.
\begin{table}[htbp!]
  \begin{center}
    \vspace{0.5cm}
    \begin{tabular}{|c|c|c|c|}\hline 
      & S & $\epsilon_{tot}$ [\%] & $B_{b\bar{b}}/S$  \\ \hline
      $B_{s}^0 \to D_s^{\mp} K^{\pm}$ & 6200 & 0.32 & $<$ 0.7  \\\hline\hline
    \end{tabular}
   \caption{\label{tab:selres}
      Summary of event yield ($S$), 
      experimental efficiency ($\epsilon_{tot}$), and background
      to signal ratios.}
  \end{center}
\end{table}

In a toy MC study multidimensional probability functions (PDFs)
are constructed. These are supposed to mimic the outcome of an analysis of data
acquired at LHCb. So first the behaviour of the experiment is described building
different PDFs. In the toy in use for the sensitivity studies of the
$B^0_s \to D_s K$ channels several PDFs describing the mass distribution,
the proper time acceptance, the flavour tagging and the particle identification response have been
built both for the signal events and for the background events according to the studies
done with the full Geant 4 simulation \cite{bib:geant}. The final PDF is built as the product of all the different PDFs and finally
a likelihood fit to the generated data is performed
to extract the parameters and their errors. 
In this toy not only the $B^0_s \to D_s K$ decay channels are considered, but
also the topologically similar $B^0_s \to D_s \pi$ decay channels.
The $B^0_s \to D_s \pi$ decay channel has a larger branching ratio and
consequently a higher annual yield (140000 events) and it allows for the
determination of the  $\Delta m_s$ parameter.
A summary of the input parameters that
are used is given in Tab.~\ref{tab:toy}.
\begin{table}[h]
  \begin{center}
    \begin{tabular}{|c|c|}\hline
       \textbf{Parameter} &  Input value \\ \hline\hline
       $\sigma(m_{B^0_s})$ (MeV) & 14 \\
       $\Delta \Gamma_s/\Gamma_s$ & 0.1\\
       $\Delta m_s (ps^{-1})$ & 17.5\\
       mistag fraction $\omega$ & 0.328 \\
       tagging efficiency $\epsilon_{tag}$ & 0.5812\\
       $|\lambda_f|$ & 0.37\\
       $\gamma + \phi_s$ ($^{\circ}$) & 60\\
       $\Delta_{T1/T2}$ ($^{\circ}$) & 0 \\
        $B^0_s \to D_s^- \pi^+$ annual yield & 140k\\
        $B^0_s \to D_s^{\mp} K^{\pm}$ annual yield & 6.2k\\
        $B^0_s \to D_s^- \pi^+$ B/S ratio & 0.2\\
        $B^0_s \to D_s^{\mp} K^{\pm}$ B/S ratio & 0.7\\ \hline\hline
    \end{tabular}
    \caption{\label{tab:toy} Input parameter values for the toy MC simulation program.}
  \end{center}
\end{table}
The final sensitivity results are shown in Tab.~\ref{tab:toyresult}. It can be seen that with one year of data
taking at nominal luminosity a sensitivity on $\gamma + \phi_s$ of 10.3$^{\circ}$ can
be obtained. A more detailed study can be found in \cite{bib:scohen}.
\begin{table}[h]
  \begin{center}
    \begin{tabular}{|c|c|c|c|c|}\hline
      &  $\Delta m_s (ps^{-1})$ &  $|\lambda_f|$ &   $\Delta_{T1/T2}$ ($^{\circ}$) &
       $\gamma + \phi_s$ ($^{\circ}$) \\ \hline\hline
Input value & 17.5 & 0.37 & 0 & 60  \\
Fitted value & 17.5 & 0.372 & 0 & 60.4  \\
$\sigma$ (1y) & 0.007 & 0.061 & 10.3 & 10.3 \\
$\sigma$ (5y) & 0.003 & 0.027 & 4.6 & 4.6 \\\hline\hline
    \end{tabular}
    \caption{\label{tab:toyresult}  Summary of the sensitivity results for the
      toy for $B^0_s \to D_s^{\mp} K^{\pm}$ and  $B^0_s \to D_s^-
      \pi^+$ events.}
  \end{center}
\end{table}

\section{Extracting $\gamma$ from $B^0 \to \pi^+ \pi^-$ and $B^0_s \to K^+K^-$}

\begin{figure}[htbp!]
  \begin{center}
    \rotatebox{270}{\includegraphics[width=0.18\textwidth]{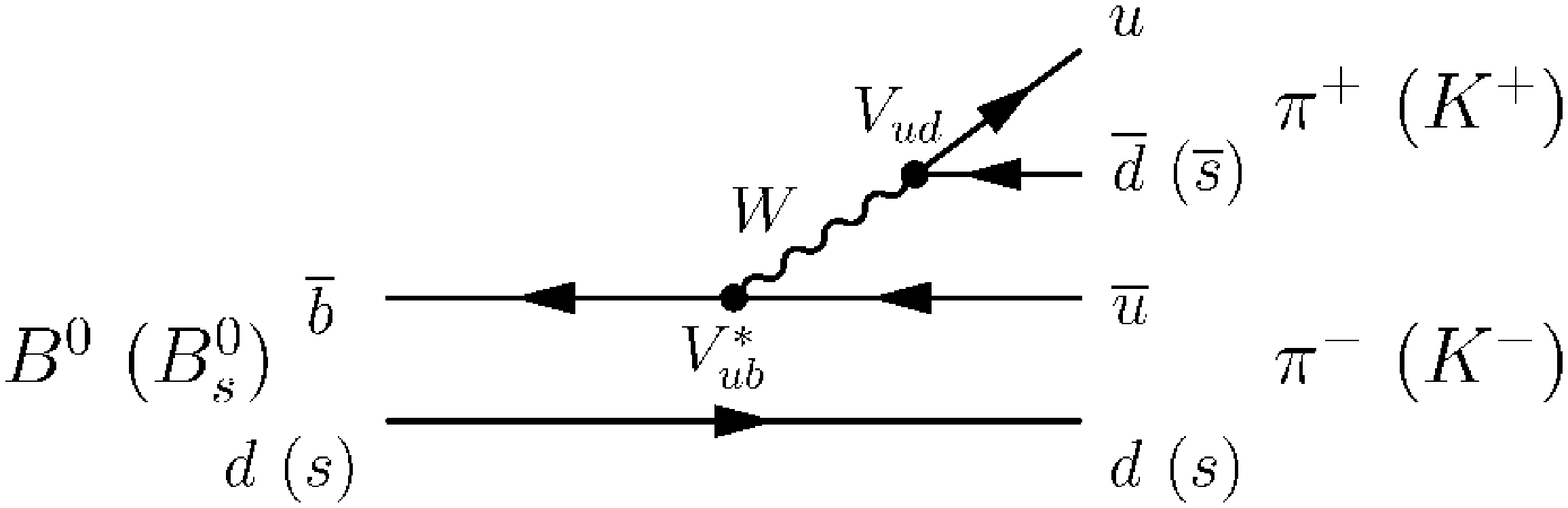}}
    \caption{\label{fig:fey} The $B_{(s)}^0 \to h^+h'^-$ tree diagram.}
  \end{center}
\end{figure}

In this section the way to extract $\gamma$ through the combined measurement of the
$B^0 \to \pi^+ \pi^-$ and $B^0_s \to K^+K^-$ CP asymmetries and under the assumption of
invariance of the strong interaction under the $d$ and $s$ quarks exchange
(U-spin symmetry)~\cite{bib:fleis} is described. 
In Fig.~\ref{fig:fey} the $B_{(s)}^0 \to h^+h'^-$ tree diagrams are shown.

For a neutral $B$-meson decaying into a CP eigenstate $f$, the time-dependent CP asymmetry
is given by:
\begin{eqnarray}
  {\cal A}_{CP}(t) &=& \frac{\Gamma(\overline{B^0}_{d/s}(t) \to f)-\Gamma(B^0_{d/s}(t) \to f)}
     {\Gamma(\overline{B^0}_{d/s}(t) \to f)+\Gamma(B^0_{d/s}(t) \to f)}  \nonumber \\  &=&
        \frac{-C_{CP}\cos\Delta mt + S_{CP}\sin\Delta mt}
             {\cosh\frac{\Delta\Gamma}{2}t - A_{CP}^{\Delta\Gamma}
               \sinh\frac{\Delta\Gamma}{2}t},
 \label{eq:cp}
\end{eqnarray}
where $\Gamma(\overline{B^0}_{d/s}(t) \to f)$ and $\Gamma(B^0_{d/s}(t) \to f)$ are the decay
rates of the initial $\overline{B}$ and $B$ states respectively, and $\Delta m$ and $\Delta \Gamma$
are the mass and width differences between the two mass eigenstates.

As shown in \cite{bib:fleis} $C_{CP}$ and $S_{CP}$ for the
$B^0 \to \pi^+ \pi^-$ and $B^0_s \to K^+K^-$ can be written as functions of 
the  Unitarity Triangle angle $\gamma$, and of the two hadronic parameters $d$ and $\theta$ (which parametrize
respectively the magnitude and phase of the penguin-to-tree amplitude ratio), as follows
\begin{eqnarray}
  \label{eq:pipi}
 C_{\pi\pi} &=& \frac{2d\sin\theta\sin\gamma}{1-2d\cos\theta\cos\gamma+d^2}\;\;, \nonumber \\
  S_{\pi\pi} &=&  \frac{\sin{(\phi_d+2\gamma}) - 2d\cos{\theta}\sin{(\phi_d+\gamma})+d^2\sin{\phi_d}}{1-2d\cos{\theta}
    \cos{\gamma}+d^2}\;\;,  \nonumber \\
  C_{KK} &=& - \frac{2d'\sin{\theta'}\sin{\gamma}}{1+2d'\cos{\theta'}\cos{\gamma}+d'^2}\;\;,  \nonumber \\
  S_{KK} &=& \frac{\sin{(\phi_s+2\gamma}) + 2d'\cos{\theta'}\sin{(\phi_s+\gamma})+
    d'^2\sin{\phi_s}}{1+2d'\cos{\theta'}\cos{\gamma}+d'^2}\;\;, \nonumber \\
  & & 
\end{eqnarray}
where $\phi_d$ and $\phi_s$ are the $B^0_d/\overline{B^0_d}$ and $B^0_s/\overline{B^0_s}$ mixing
phases which in LHCb will be measured from $B_{d}^{0} \to J/\psi K_{S}$ decay and from
$B_{s}^0 \to J/\psi \phi$ respectively~\cite{bib:lfernandez}. This system of four equations
and five unknowns ($d,d',\theta,\theta'$ and $\gamma$) can be solved with the help of
the U-spin symmetry, 
as a consequence  $ d = d' $ and $\theta = \theta' $. This results in an over-constrained
system of three unknowns and four equations.

\subsection{Sensitivity to the CP asymmetries and to $\gamma$}

As a result of the event selection, the annual yields together with the total efficiencies and with the background to signal
ratios are listed in Table~\ref{tab:sel2}.

\begin{table}[htbp!]
  \begin{center}
    \vspace{0.5cm}
    \begin{tabular}{|c|c|c|c|c|c|} \hline
      Event type & $BR$ & S & $\epsilon_{tot}$ [\%] &
      $B_{b\bar{b}}/S$ & $B_{sp}/S$ \\ 
      &  ($\times 10^{-6}$) & & & & \\ \hline\hline
      $B^{0}_d \to \pi^+\pi^- $ & 4.8 & 35700 & 0.93 & 0.46 & 0.08 \\
      $B^{0}_d \to K^+\pi^- $ &18.5 & 137600 &  0.93 & 0.14 & 0.02\\
      $B^{0}_s \to \pi^+K^- $ & 4.8 & 9800 &  1.02 & 1.92 & 0.54\\
      $B^{0}_s \to K^+K^- $ & 18.5 & 35900 &  0.97 & $<$ 0.06 & 0.06\\ \hline\hline
    \end{tabular}
   \caption{\label{tab:sel2}
      Summary of event yields ($S$), Branching ratios,
      experimental efficiency ($\epsilon_{tot}$), and background
      to signal ratios, considering both the $b\bar{b}$ combinatorial
      background and the specific background. }
  \end{center}
\end{table}

As for the  $B^0_s \to D_s K$ decays a toy MC is used to mimic the outcome of an analysis of data
acquired at LHCb. In this toy, used for the CP sensitivity studies of the
$B_{(s)}^0 \to h^+h'^-$ channels, several PDFs describing the mass distribution,
the proper time acceptance, the flavour tagging and the particle ID response have been
built both for the signal events and for the background events according to the studies
done with the full Geant 4 simulation~\cite{bib:sens}.
The final PDF is built as the product of all the different PDFs and 
a likelihood fit to the generated data is performed
to extract the parameters and their errors.

The extraction of $\gamma$ is then performed using a Bayesian approach in three different
U-spin scenarios:
\begin{itemize}
\item Assuming perfect U-spin symmetry: $d = d'$ and $\theta = \theta'$ .
\item With a weaker assumption on the U-spin symmetry: $d = d'$
and no constraint on $\theta$ and $\theta'$.
\item With an even weaker assumption on the U-spin symmetry: $\xi = d'/d =$ [0.8,1.2]
and no constraint on $\theta$ and $\theta'$.
\end{itemize}
In the three different cases the sensitivity on
$\gamma$ is taken considering the 68\% probability interval of the resulting
PDF distribution for $\gamma$.

\begin{figure}[htbp!]
  \begin{center}
    \includegraphics[width=0.35\textwidth]{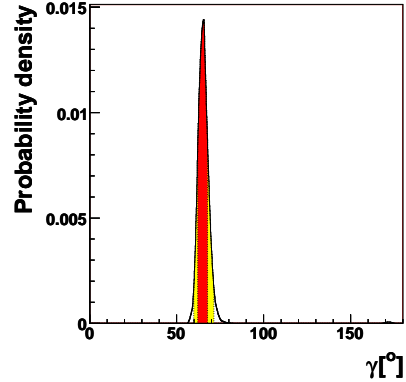}
    \caption{\label{fig:gamma} $\gamma$ probability density obtained
      in case of perfect U-spin symmetry.
      The 68\% and 95\%  probability intervals are visible.
      The 68\%  probability interval corresponds to a sensitivity of 4$^{\circ}$.
      The plot has been obtained using tools developed by the UTfit
      Collaboration~\cite{bib:utfit}.}
  \end{center}
\end{figure}

Fig.~\ref{fig:gamma} shows an example of a resulting PDF for $\gamma$ obtained assuming perfect U-spin symmetry.
The 68\% and 95\% probability intervals are visible.
The 68\% probability interval corresponds to a sensitivity of 4$^{\circ}$.
The sensitivity in the second and in the third scenarios increases and varies
between 7$^{\circ}$ and 10$^{\circ}$.

It is important to note that the extraction of $\gamma$ by means of the  $B^0 \to \pi^+ \pi^-$ and $B^0_s \to
K^+K^-$ decays uses not only tree diagrams but also loop diagrams and is
therefore sensitive to new physics.

\section{Conclusions}
In these proceedings three complementary methods for the extraction of 
the Unitarity Triangle  angle $\gamma$ have been discussed.

The potential of the $B \to DK^{(*)}$ decays has been studied by employing the
combined Gronau-London-Wyler (GLW) and the Atwood-Dunietz-Soni (ADS) methods.
For the charged $B$-meson, with one year of data at nominal luminosity, 
the angle $\gamma$ can be determined with a precision in the range
$8^{\circ}-10^{\circ}$, depending on the value of the strong 
phase $\delta_D$ \cite{bib:patel}. For the neutral $B$, the angle $\gamma$ 
can be determined with a precision smaller than $10^{\circ}$, depending on 
the value of the strong phase $\delta_B$ and for $r_B$ values bigger 
than 0.3 \cite{bib:kazu}. 

It has been shown that the angle $\gamma$ can also be extracted
with a time-dependent analysis through the $B_s \to D_s K$ decays,
provided that the $B_s$ mixing phase is measured independently.
With one year of data taking at nominal luminosity,
a sensitivity on $\gamma + \phi_s$ of 10$^{\circ}$ can be obtained.

The combined measurement of the $B^0 \to \pi^+ \pi^-$ and $B^0_s \to K^+K^-$
time-dependent CP asymmetries allows the determination of the Unitarity Triangle angle $\gamma$,
up to U-spin flavour symmetry breaking corrections. Here a sensitivity of
10$^{\circ}$, with one year of data taking at nominal luminosity, can also be
obtained, but the final results depend on the assumption on the
breaking of the U-spin symmetry. This method uses not only tree diagrams but also loop diagrams and is
therefore sensitive to new physics. 

Other methods, including $B \to D (KK\pi\pi)K$ and $B \to D (K_s\pi^+\pi^-)K$
decays, to extract $\gamma$ at LHCb have been studied. More
details can be found in \cite{bib:15,bib:13}.

\end{document}